\documentclass[twocolumn,superscriptaddress,floatfix,preprintnumbers,prl ,nofootinbib,hyperref]{revtex4-2}
\usepackage[colorlinks=true,breaklinks=true]{hyperref}
\usepackage[utf8]{inputenc}
\hypersetup{allcolors=[rgb]{0.0 0.0 0.6},linkcolor=[rgb]{0.75 0.05 0.05}}
\usepackage{mathtools}
\usepackage[dvipsnames]{xcolor}
\usepackage{float}
\usepackage{letltxmacro}
\LetLtxMacro{\oldcite}{\cite}
\renewcommand{\cite}[1]{\mbox{\oldcite{#1}}}

\usepackage{orcidlink}

\long\def\exclude#1{}

\DeclareMathOperator{\GeV}{GeV}

\DeclareMathOperator{\MeV}{MeV}

\newcommand{\beq}{\begin{equation}}
\newcommand{\eeq}{\end{equation}}

\def\ga{\,\,\raise0.14em\hbox{$>$}\kern-0.76em\lower0.28em\hbox
{$\sim$}\,\,}

\newcommand{\wP}{\omega_{\rm p}}
\newcommand{\kS}{k_{\rm s}}

\newcommand{\RNS}{R_{\rm NS}}

\allowdisplaybreaks

\bibpunct{[}{]}{,}{n}{}{,}

\begin{document}

\title{Low-Energy Supernovae Severely Constrain Radiative Particle Decays}

\author{Andrea Caputo \orcidlink{0000-0003-1122-6606}} 
\affiliation{School of Physics and Astronomy, Tel-Aviv University, Tel-Aviv 69978, Israel}
\affiliation{Department of Particle Physics and Astrophysics, Weizmann Institute of Science, Rehovot 7610001, Israel}

\author{Hans-Thomas Janka \orcidlink{0000-0002-0831-3330}}
\affiliation{Max-Planck-Institut f\"ur Astrophysik, Karl-Schwarzschild-Str.~1, 
85748 Garching, Germany}

\author{Georg Raffelt \orcidlink{0000-0002-0199-9560}} 
\affiliation{Max-Planck-Institut f\"ur Physik (Werner-Heisenberg-Institut), F\"ohringer Ring 6, 
80805 M\"unchen, Germany}

\author{Edoardo Vitagliano \orcidlink{0000-0001-7847-1281}} 
\affiliation{Department of Physics and Astronomy, University of California, Los Angeles, California 90095-1547, USA}



\begin{abstract}

The hot and dense core formed in the collapse of a massive star is a powerful source of hypothetical feebly-interacting particles such as sterile neutrinos, dark photons, axion-like particles (ALPs), and others. Radiative decays such as $a\to2\gamma$ deposit this energy in the surrounding material if the mean free path is less than the radius of the progenitor star. For the first time, we use a supernova (SN) population with particularly low explosion energies as the most sensitive calorimeters to constrain this possibility. These SNe are observationally identified as low-luminosity events with low ejecta velocities and low masses of ejected $^{56}$Ni. Their low energies limit the energy deposition from particle decays to less than about 0.1~B, where $1~{\rm B~(bethe)}=10^{51}~{\rm erg}$. For 1--500~MeV-mass ALPs, this generic argument excludes ALP-photon couplings $G_{a\gamma\gamma}$ in the $10^{-10}$--$10^{-8}~{\rm GeV}^{-1}$ range.

\end{abstract}

\maketitle


\textit{Introduction.}---Collapsed stars are powerful astrophysical factories of neutrinos and, possibly, new feebly interacting particles such as sterile neutrinos, dark photons, new scalars, axions and axion-like particles (ALPs), and many others~\cite{Raffelt:1996wa,Raffelt:2006cw}. Unique constraints on their interaction strengths derive from the cooling speed of the proto-neutron star (PNS) formed in the historical supernova (SN) 1987A, estimated from the neutrino signal to be a few seconds. Similar limits, with different systematic uncertainties, come from neutron-star cooling times.

Despite their feeble interactions, new particles can become visible by radiative decays. The temperature scale of a SN core is 30~MeV, resulting in an energy scale of 100~MeV for the emitted particles and their decay products. The decay photons from all past SNe contribute excessively to the cosmic diffuse $\gamma$-ray background unless the energy loss of an average SN in this form is below 0.015--0.03~B, using (0.5--$1)\times10^{7}~{\rm Mpc}^{-3}$ for the comoving cosmic density of all past core collapses \cite{Cowsik:1977vz,DeRocco:2019njg,Calore:2020tjw,Calore:2021klc,Caputo:2021rux}. This small energy loss is to be compared with a typical neutron-star binding energy of 200--400~B, depending on its mass and the nuclear equation of state. 

Here it was assumed that the mean free path (MFP) against radiative decay is less than about the Hubble scale. For yet faster decays, even more restrictive limits arise from the absence of excess events in the Gamma-Ray Spectrometer on board the Solar Maximum Mission satellite that operated at the time of the SN~1987A neutrino detection \cite{Chupp:1989kx,Kolb:1988pe,Oberauer:1993yr,Jaeckel:2017tud,DeRocco:2019njg}.

These arguments are moot if the decay is so fast that most of the electromagnetic energy is dumped within the progenitor star (radius 3--$50\times10^{12}$~cm for Type-II SNe) surrounding the collapsing core. This energy deposition contributes to the SN explosion energy, often taken to be 1--2~B, thus providing a ``calorimetric'' constraint on radiative decays that can also include the $e^+e^-$ channel. This idea was first advanced by Falk and Schramm a decade before SN~1987A \cite{Falk:1978kf}, recently rediscovered \cite{Sung:2019xie}, and applied to muon-philic bosons \cite{Caputo:2021rux} and generic $e^+e^-$ decays \cite{Calore:2021lih}. Moreover, it was speculated that such effects could power SN  explosions~\cite{1982ApJ...260..868S,Rembiasz:2018lok,Mori:2021pcv} or gamma-ray bursts~\cite{Berezhiani:1999qh,Diamond:2021ekg}.

In this \textit{Letter} we argue that such scenarios are strongly constrained because, instead of considering a ``typical'' explosion energy to rule such effects in or out, one should use the lowest-energy well-established cases. For example, reconstruction of the explosion energy of SN~1054 that has led to the Crab Nebula and its pulsar suggests a value 0.1~B or less \cite{Yang:2015ooa,Stockinger+2020}.

\textit{Low-energy Supernovae (LESNe).}---Other than this particular case, there is an entire class of core-collapse SNe (CCSNe) with similar low explosion energies. Observationally, CCSN energies and luminosities cover wide ranges. The luminosity during the light-curve plateau and tail phases of Type II-P events is tightly correlated with the ejected mass including explosively nucleosynthesized $^{56}$Ni and with the explosion energy (see the recent compilations \cite{Spiro+2014,Pejcha:2015pca,Muller:2017bdf,Pejcha2020,Yang+2021}). The radioactive decays of $^{56}$Co, the daughter nucleus of $^{56}$Ni, to stable $^{56}$Fe heat the expanding SN debris and thus increase the luminosity and extend the duration of the plateau (e.g., \cite{Goldberg+2019,Kozyreva+2019}). Of particular interest for our study is the subgroup of \hbox{LESNe}, which exhibit $^{56}$Ni masses of only several $10^{-3}\,M_\odot$, over ten times less than the average of most CCSNe \cite{Pastorello+2004}. These SNe are also 10--100 times dimmer than normal CCSNe, and their 2--3 times lower photospheric expansion velocities point to minimal explosion energies around 0.1\,B or even less (e.g., \cite{Spiro+2014,Valenti+2009}).  

The relative contribution of such LESNe could be several percent of all CCSNe \cite{Pastorello+2004}, but it might also be considerably higher because of faintness-related observational selection effects and possible misclassification of dim, low-energy events \cite{Valenti+2009}. Hydrodynamical modeling of explosions and observables suggests low-mass or moderate-mass red supergiant (RSG) stars as origins of LESNe of Type II-P \cite{Pastorello+2009,Spiro+2014,Lisakov:2017uue,Yang+2021}, possibly including electron capture SNe from the collapse of super-asymptotic giant-branch stars with oxygen-neon-magnesium cores instead of iron cores (e.g., \cite{Pumo+2009,Hiramatsu+2021}).

These observational findings are compatible with predictions from self-consistent simulations of neutrino-driven explosions for low-mass Fe-core and O-Ne-Mg-core progenitors ($M \lesssim 12\,M_\odot$) \cite{Kitaura+2006,Janka+2008,Huedepohl+2010,Fischer+2010,Melson+2015,Radice+2017,Burrows+2019,Glas+2019,Mueller+2019,Stockinger+2020,Zha+2021} and corresponding determination of multi-band light-curve properties \cite{Kozyreva+2021} and nebular spectra \cite{Jerkstrand:2017hbi}. Explosions by neutrino heating occur after only a short phase of shock-wave stagnation, form low-mass neutron stars (NSs) with baryonic masses $\lesssim$\,$1.36\,M_\odot$, and develop explosion energies around 0.1\,B. 

Therefore, observations of LESNe, in line with state-of-the-art, self-consistent models, suggest that the additional energy deposition by radiative particle decays should be significantly below about 0.1\,B.

\textit{Dynamical effects of energy deposition.}---Observation\-ally determined ejecta masses of low-luminosity, LESNe of Type II-P suggest low-mass
RSG progenitors with $M\simeq 9$--$15\,M_\odot$ \cite{Pastorello+2009,Spiro+2014,Yang+2021} and radii between $R_*\simeq 5\times 10^{12}$ and several $10^{13}$\,cm. Such values agree with stellar evolution models (see \cite{Stockinger+2020} for a few examples) and with the direct observation of the RSG progenitor of the low-luminosity SN~2008bk, whose radius was $(3.45\pm 0.24)\times 10^{13}$\,cm at 6~months prior to explosion \cite{VanDyk+2012}. On the other hand, the progenitor of SN~1987A, the blue supergiant star Sanduleak $-69^\circ 202$, had only $R_* = (3\pm1)\times10^{12}\,\mathrm{cm}$~\cite{1988ApJ...330..218W,Menon+2017,Utrobin+2021}. 

Let us now consider new particles $a$ that emerge from the SN core for, say, 3\,s and propagate nearly with the speed of light $c$. This pulse with
length $\Delta R_a \simeq c \times 3\,\mathrm{s} \sim 10^{11}\,\mathrm{cm}$ and average luminosity $L_a^i = N_a E_a c/(3\,\mathrm{s})$ sweeps through the star and reaches the surface only some 100\,s later. ($N_a$ is the total number of initially created particles and $E_a$ their average energy.) Particle decays imply $L_a(R) = L_a^i\exp{(-R/\lambda_a)}$ at radius $R$, where in particular we consider the MFP range $\RNS< \lambda_a < R_\ast$ with a PNS radius
$\RNS \simeq 2\times 10^6$\,cm and $R_\ast \simeq (3$--$50)\times 10^{12}$\,cm, bracketing the range discussed earlier.

We consider ``radiative'' decays $a\to\gamma\gamma$ or $e^+e^-$ that quickly thermalize in the medium. For the largest $m_a$ of a few $100\,\MeV$, even $a\to\mu^+\mu^-$ or $\pi^+\pi^-$ is kinematically possible, but then some decay energy is lost in neutrinos. $100\,\MeV$ photons are mainly absorbed by pair production on nuclei, so the $a\to\gamma\gamma$ and $e^+e^-$ channels are essentially equivalent. The mass attenuation coefficient in hydrogen for $E_\gamma=100\,\MeV$ is around $100\,{\rm g}/{\rm cm}^2$ \cite{Groom:2020}, for $\rho=10^{-8}\,{\rm g}/{\rm cm}^3$ near the progenitor surface implying  a MFP of $10^{10}\,{\rm cm}$, much smaller than $R_*$ and $\Delta R_a$.

The detailed dynamical impact of this energy deposition is subtle, because the importance of momentum transfer and thus the relative contributions of kinetic and thermal energies are determined by the local ratio of $L_a(R)$ and the Eddington luminosity limit for the decaying $a$'s. The latter is the critical $L_a$ where the momentum force equals the gravitational one and is $L_\mathrm{Edd}(R) = 4\pi GcM(R)\rho(R)\lambda_a \simeq 0.5\,{\rm B}\,{\rm s}^{-1}
[M(R)/M_\odot]\rho \lambda_{13}$, where $M(R)$ is the mass enclosed by radius $R$, $\rho(R)$ the local matter density in g\,cm$^{-3}$, and $\lambda_{13} = \lambda_a/(10^{13}\,\mathrm{cm})$. If $L_a(R)\simeq L_\mathrm{Edd}(R)$, particle decays directly accelerate the medium, whereas if $L_a(R)\ll L_\mathrm{Edd}(R)$ the energy deposition mostly heats the stellar gas. This thermal energy is converted to kinetic energy of expansion by hydrodynamic $pdV$ work. In any case and independently of the detailed processes, however, the entire decay energy is thus dumped into the progenitor star if $\lambda_a < R_\ast$. 

The gravitational binding energy of all layers outside the PNS in low-mass progenitors is at most some 0.01\,B (see, e.g., \cite{Stockinger+2020}) and orders of magnitudes less in the He shell and H envelope. Therefore, decay energy deposition of $>$\,0.1\,B can cause powerful ejection of the stellar material and will gravitationally unbind most of the progenitor mass, independently of neutrino heating or any other hypothetical explosion mechanism. Of course, such particle decay could not explain the explosion of ``normal'' CCSNe that have much larger energies.

Momentum and energy deposition is locally quasi instantaneous, because the time scale $\Delta R_a/c$ of the $a$ pulse is much shorter than the local sound travel time $\Delta R_a/c_\mathrm{s}(R)$, where $c_\mathrm{s}(R)$ is the sound speed. It is $c_\mathrm{s}\simeq c/30$ near the PNS and decreases with $R$. Therefore, the sudden input of huge amounts of momentum and thermal energy creates a pressure wave that steepens into an outgoing shock. This alone will cause a SN-like outburst or it will strengthen the blast wave powered by the SN mechanism, which leaves the collapsed stellar core only later but ultimately will merge with the $a$-driven ejecta.

The energy injected between the PNS and the progenitor surface is
\begin{multline}
    E_{\rm mantle} = \int dt \int_0^{\RNS} dR \int_{m_a'(R)}^\infty d E_a \, \,\frac{dL_a(R,E_a,t)}{dR\,d E_a}\\ 
    \times \bigl\{\exp[-(\RNS- R)/\lambda_a] - \exp[-(R_\ast - R)/\lambda_a]\bigr\},
\label{Eq:mantle}
\end{multline}
where $dL_a(R,E_a,t)/dR\,d E_a$ is the differential change of $L_a$ over distance $dR$ and energy $d E_a$ at radius $R$ and energy $E_a$. In the exponentials we assume radial propagation, causing a geometric error close to the PNS. This effect matters only if $\lambda_a\simeq\RNS$, where our argument is only approximate anyway (see below). The lower $dE_a$ integration limit $m_a'(R)$ is chosen to exclude gravitationally bound particles, and we account for energy redshifting of the escaping ones, although such corrections are not shown in Eq.~(\ref{Eq:mantle}) (see Supplemental Material). The time integration is performed over all available time snapshots of our numerical SN model, although the bulk of the emission lasts only for $\sim$3\,s. The radial integration extends over the PNS with $\RNS \simeq 20$\,km. The difference of the exponential functions with energy-dependent MFP $\lambda_a(E_a')$ (redshifted energy $E_a'$) accounts for the cumulative energy deposition by particle decays between $\RNS$ and $R_\ast$, considering that $L_a(R)\{1-\exp[-(R_1-R)/\lambda_a]\}$ is the luminosity produced at $R$ that has decayed before reaching $R_1$.

\textit{Reference SN model for particle emission.}---To apply our argument we need to model particle emission from the SN core. As a reference case it is convenient to use the Garching group's muonic model SFHo-18.8 \cite{Bollig:2020xdr,CCSNarchive} that we employed earlier to study muon-philic boson emission \cite{Bollig:2020xdr,Caputo:2021rux} (see also \cite{Croon:2020lrf}). This is the coldest of the Garching muonic models and has a peak temperature of 30--$40\,\MeV$ and final NS mass of $1.351\,M_\odot$ (baryonic) and $1.241\,M_\odot$ (gravitational), implying a binding energy of $0.110\,M_\odot=197\,{\rm B}$. 

These properties are compatible with predictions by theoretical models for LESNe \cite{Kitaura+2006,Janka+2008,Huedepohl+2010,Fischer+2010,Melson+2015,Radice+2017,Burrows+2019,Glas+2019,Mueller+2019,Stockinger+2020,Zha+2021}. Such a small mass is close to the minimum NS mass expected to be formed in CCSNe. Therefore, our analysis is on the conservative side, because the particle production is weaker than in more massive and hotter PNSs. Of course, the stellar progenitor model and collapse and explosion dynamics of SFHo-18.8 is not representative for LESNe, but the core is a good proxy for our needs in that it is compatible with all constraints on the nuclear equation of state and includes muons, a physically unavoidable feature.

For comparison, we have also considered an $8.8\,M_\odot$ electron-capture SN \cite{Huedepohl+2010} with a baryonic NS mass of $1.366\,M_\odot$ and binding energy of $166\,{\rm B}$ that used Shen's nuclear equation of state and also reaches peak temperatures of 30--$40\,\MeV$, but does not include muons and PNS convection. We find rather similar results, which demonstrates that our particle bounds are robust and to a large extent not sensitive to many ingredients of the PNS cooling models. For our argument primarily the mass of the NS and its peak temperature and average density are crucial, a fact that will be consolidated by a simple one-zone model discussed later.

The small allowed amount of particle emission is a negligible perturbation, in contrast to the traditional SN cooling argument, so here it is self-consistent to use an unperturbed reference model. The total particle emission depends primarily on the temperature reached within the SN core and for how long it stays hot. Again our main reference model is conservative because it cools quickly due to convection (which is numerically implemented with a mixing-length treatment, as detailed in Refs.~\cite{Mirizzi:2015eza,Huedepohl2014}). Another model used in the literature to estimate the total ALP emission from SN~1987A \cite{Payez:2014xsa} has an emission period around 3~times longer (see Ref.~\cite{Caputo:2021rux} for an explicit comparison).

\textit{ALP production and decay.}---Our general argument pertains to any new mechanism for energy transfer from the inner SN core to the progenitor star, but as a specific example we study ALPs, pseudoscalar bosons that interact exclusively through a two-photon channel given by 
$\mathcal{L}_{a\gamma\gamma}=G_{a\gamma\gamma}a\, {\bf E}\cdot{\bf B}$, where $G_{a\gamma\gamma}$ is a coupling constant of dimension (energy)$^{-1}$.
(Henceforth we use natural units with $\hbar=c=1$.) One key process is Primakoff production on charged particles
$\gamma+Ze\to Ze+a$ by photon exchange \cite{Dicus:1979ch,Raffelt:1985nk,DiLella:2000dn,Lucente:2020whw} with a cross section $\sigma_{\rm P}=\frac{1}{2}Z^2\alpha G_{a\gamma\gamma}^2f_{\rm P}$, where $f_{\rm P}\simeq 1$ depends on the screening scale, the plasma frequency, and the ALP mass. For the conditions of a SN core, the energy emission rate per unit volume is
\begin{equation}
    Q_{\rm P}\simeq \hat{n}\,\frac{2\alpha G_{a\gamma\gamma}^2}{3\pi^2}\,
    \bigl(m_a^2+3m_aT+3T^2\bigr)T^2 e^{-m_a/T}\,,
\end{equation}
where $\hat n\simeq (1-Y_n)n_B$ is the effective charged-particle density, $Y_n$ the neutron number per baryon, and $n_B$ the baryon density (see Supplemental Material for details).

For $m_a\agt T$ this rate quickly drops with increasing mass, whereas photon coalescence $\gamma + \gamma \rightarrow a$ becomes important \cite{Lucente:2020whw}. The corresponding energy-loss rate is 
\begin{equation}
    Q_{\gamma\gamma\to a}=\frac{G_{a\gamma\gamma}^2T^3m_a^4}{128\pi^3}\,F(m_a/T)\,.
\end{equation}
Previously this rate was calculated with Maxwell-Boltz\-mann (MB) statistics \cite{DiLella:2000dn,Lucente:2020whw}. In this case $F_{\rm MB}(\mu)=\mu^2 K_2(\mu)$, where $K_2(\mu)$ is a modified Bessel function of the second kind. For small arguments it is $2/\mu^2$, for large arguments $e^{-\mu}\sqrt{\pi/2\mu}$. We have also derived the full expression for Bose-Einstein statistics (see Supplemental Material), providing a somewhat larger $Q_{\gamma\gamma\to a}$. Coalescence becomes more important than Primakoff for $m_a\agt70\,\MeV$ and reaches a maximum near $m_a\simeq200\,\MeV$.

ALPs decay by the reverse process $a\to\gamma\gamma$ with the rate $\Gamma_{a\to\gamma\gamma}=G_{a\gamma\gamma}^2 m_a^3/64\pi$. Including a Lorentz factor for the decay rate and a velocity factor to convert the lifetime into a MFP against decay, we find
\begin{equation}\label{Eq:mfp}
    \lambda_{a\to 2\gamma}=\frac{64\pi}{G_{a\gamma\gamma}^2}\,\frac{\sqrt{E_a^2-m_a^2}}{m_a^4}
    \simeq\frac{4.0\times10^{13}\,{\rm cm}}{G_9^2}\,\frac{E_{100}}{m_{10}^4}\,,
\end{equation}
where $G_9=G_{a\gamma\gamma}/10^{-9}\,\GeV^{-1}$, $E_{100}=E_a/100\,\MeV$ and $m_{10}=m_a/10\,\MeV$ and the second equation refers to the relativistic case $E_a\gg m_a$.

\begin{figure}[b!]
\vskip-8pt
\centering
\includegraphics[width=0.9\columnwidth]{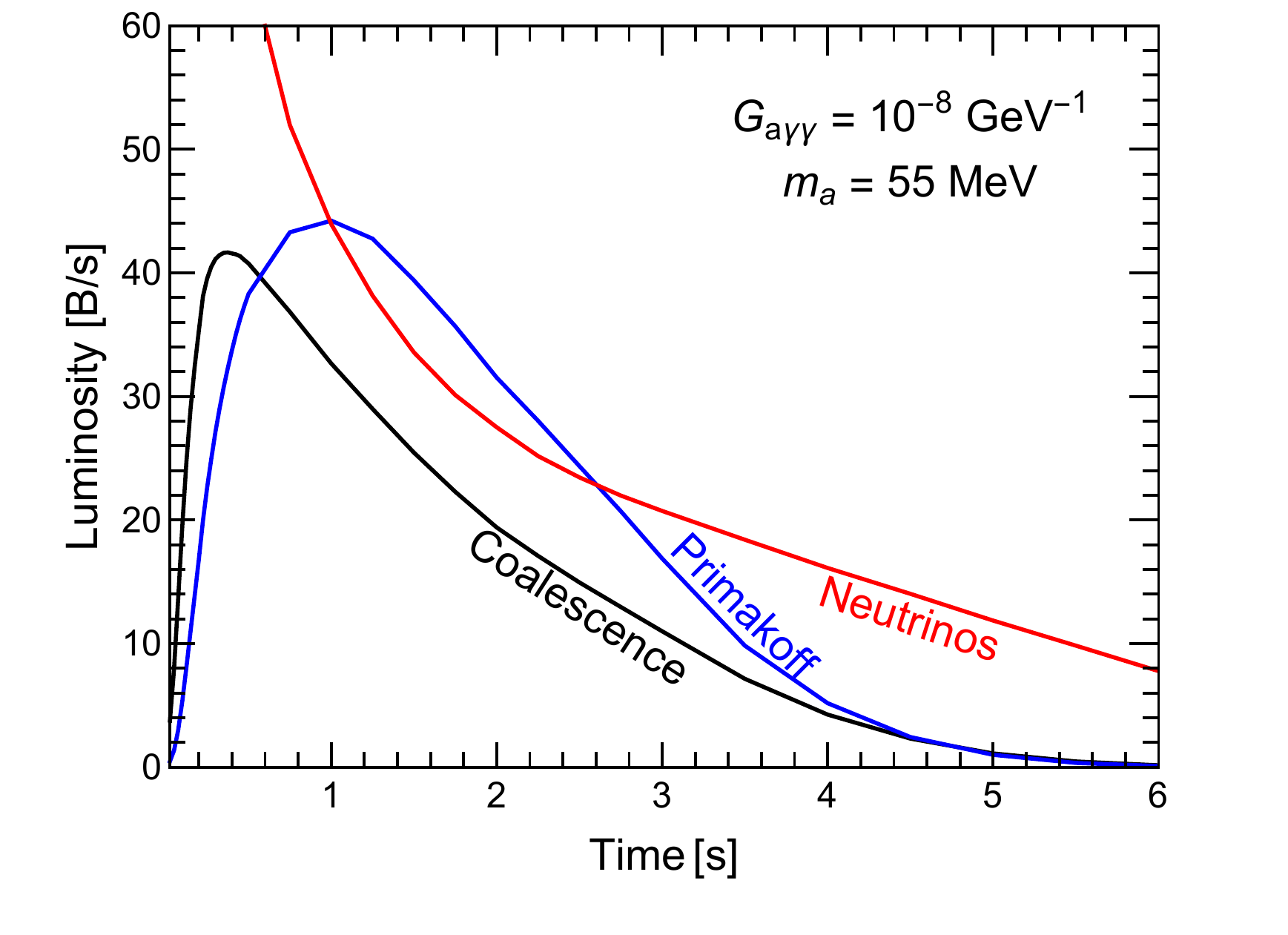}
\vskip-16pt
\caption{Luminosity evolution of our reference SN model for neutrinos (red), ALPs by Primakoff emission (blue) and by photon coalescence (black). The ALP mass $m_a=55\,\MeV$ was chosen to make Primakoff and coalescence roughly equal. $G_{a\gamma\gamma} = 10^{-8}\,\GeV^{-1}$ was chosen to match $L_\nu$ at 1\,s post bounce, corresponding to the SN~1987A cooling argument.}\label{fig:LuminosityProfile}
\vskip-3pt
\end{figure}

\textit{ALP constraints.}---Implementing ALP production in our reference model, we find the luminosity evolution shown in Fig.~\ref{fig:LuminosityProfile}, where $m_a=55\,\MeV$ was chosen such that Primakoff emission and photon coalescence are comparable. For smaller $m_a$, Primakoff dominates, for larger $m_a$ it is coalescence. As the temperature profile evolves from initially highest $T$ in the hot accretion mantle of the PNS to a $T$-maximum in its high-density core seconds later (see, e.g., \cite{Burrows:1986}), the relative importance of Primakoff emission and photon coalescence changes with time. Primakoff emission, which increases with $T$ and $n_B = \rho/m_B$ ($m_B$ is the nucleon mass), peaks only around 1\,s, long after core bounce, because it becomes fully effective only when the PNS has heated up in its high-density core. A similar behaviour is expected for any process that depends both on $T$ and $\rho$. Photon coalescence shows a different evolution in that it peaks much earlier, because it depends only on $T$ and therefore this process is already effective in the hot PNS mantle shortly after bounce.

All luminosities are understood for a distant observer, i.e., we implement redshift corrections for particles emitted deeply in the PNS gravitational potential and, for larger $m_a$, we discard those that are gravitationally trapped (for details, see Supplemental Material). 

\begin{figure}[ht]
\vskip-6pt
\centering
\includegraphics[width=0.9\columnwidth]{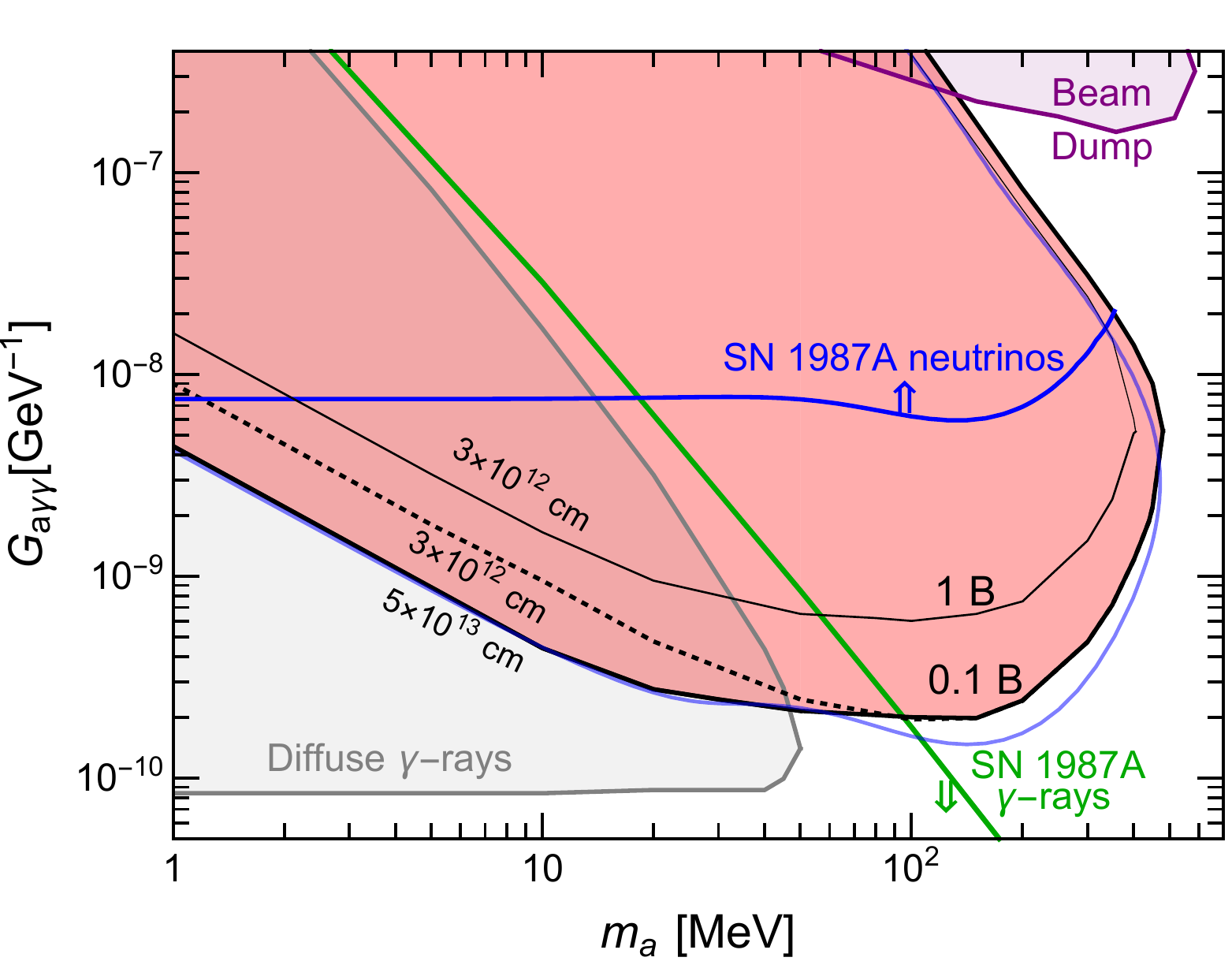}
\vskip-6pt
\caption{ALP parameters ruled out by radiative energy deposition $<0.1$\,B and progenitor radius $R_\ast = 5 \times 10^{13}$\,cm (red-shaded region). The thin blue line uses a one-zone PNS description instead of the Garching reference model (see text for details). The dotted line uses a smaller progenitor with $R_\ast = 3\times 10^{12}$\,cm, whereas the thin solid line in addition relaxes the constraint to 1\,B energy deposition. We also show the indicated constraints from beam dump experiments~\cite{Dolan2017}, the extragalactic $\gamma$-ray background, $\gamma$-ray observations of SN~1987A, and the SN~1987A neutrino signal \cite{Caputo:2021rux}. Here the double arrows point towards the excluded regions.}\label{fig:Bounds}
\vskip-6pt
\end{figure}

We finally compute the time-integrated energy deposition $E_{\rm mantle}$ (Eq.~\ref{Eq:mantle}) 
and show in Fig.~\ref{fig:Bounds} our reference limit for $E_{\rm mantle}<0.1$\,B and a progenitor with $R_\ast = 5 \times 10^{13} \, \rm cm$ (red-shaded). The dotted line uses a smaller star ($R_\ast = 3\times 10^{12}$\,cm), whereas the thin solid line in addition relaxes the constraint to $E_{\rm mantle}<1$\,B, although this weaker case is only shown for illustration.

The lower parts of the curves for small masses obey $G_{a\gamma\gamma} \propto m_a^{-1}$. For these parameters, $\lambda_{a\to2\gamma}$ is large, so the exponentials in Eq.~\eqref{Eq:mantle} can be expanded and $E_{\rm mantle} \propto G_{a\gamma\gamma}^2 \times R_\ast\, G_{a\gamma\gamma}^2 m_a^4 $. The first factor comes from ALP production, the second one from decay. 

The upper parts of the curves instead correspond to $\lambda_{a\to2\gamma}\simeq \RNS$, but should not be taken as rigorous results. Here ALPs dominate energy transfer within the PNS and probably also deposit too much energy outside, but this ``trapping regime'' requires a more detailed study to make the contours precise. Nevertheless, for MFPs larger than $\RNS$ and smaller than a few times this radius, the energy deposition is huge and therefore excluded by LESNe despite our approximations.

Finally, for $\lambda_{a\to2\gamma}\agt R_*$ the ALPs decay outside of the progenitor and contribute to the diffuse cosmic $\gamma$-ray background~\cite{Cowsik:1977vz}. Extending the recipe described in Ref.~\cite{Caputo:2021rux} to larger ALP masses we find the gray-shaded exclusion range
shown in Fig.~\ref{fig:Bounds}. We also indicate the constraint from the absence of excess $\gamma$-ray counts in conjunction with the SN~1987A neutrino signal \cite{Jaeckel:2017tud} (solid green), the bounds from beam dump experiments~\cite{Dolan2017} (purple contour), and the free streaming limit of the neutrino cooling argument for SN1987A~\cite{Caputo:2021rux} (solid blue).  

\textit{One-zone SN core.}---For a quick estimate of the excluded parameters of more general cases beyond ALPs, one can use a schematic one-zone model of the SN core. In analogy to a simple method to estimate SN~1987A cooling limits~\cite{Raffelt:2006cw}, we propose to use $T=30\,\MeV$ and nuclear density $\rho=3\times10^{14}\,{\rm g}\,{\rm cm}^{-3}$, corresponding to a baryon density of $0.181\,{\rm fm}^{-3}$. A baryonic NS mass of $1.35\,M_\odot$ implies a volume of $9000\,{\rm km}^3$ and a core radius of $12.9\,{\rm km}$. Assuming a proton abundance for Primakoff emission of $Y_p=0.15$, a cooling duration of $3\,{\rm s}$, and $R_\ast = 5 \times 10^{13}\, \rm cm$ we find the thin-blue exclusion contour in Fig.~\ref{fig:Bounds}. While our parameters were somewhat calibrated to achieve good agreement, the main features of this plot follow from overall properties of the SN core.

\textit{Discussion and outlook.}---We have argued that the low explosion energies observed in certain low-luminosity CCSNe constrain the total energy deposition in the progenitor star by radiative particle decays to less than about 0.1\,B. Specifically for ALPs, an otherwise allowed range of $m_a$ and $G_{a\gamma\gamma}$ is ruled out (see~Fig.~\ref{fig:Bounds}). 

Instead of using a time sequence of numerical PNS models, the main results also follow from a schematic one-zone PNS representation that may be useful for a first exploration of more general cases beyond ALPs.

We have {\em not} assumed that the usual neutrino-driven mechanism powers CCSNe, although neutrino heating exterior to the PNS is unavoidable. In 9--$10\,M_\odot$ progenitors, energy transfer of $\ga$0.1\,B by particle decays alone could drive SN-like mass ejection with sufficient energies. If neutrino heating does not trigger the explosion, ongoing accretion and the growing PNS mass imply higher $T$, enhanced particle emission, and thus boosted radiative decay energy deposition in the overlying star, making our argument more conservative.

The very restrictive limits from LESNe suggest that such effects cannot play a major role for the much larger explosion energies of ``normal'' CCSNe, a conclusion previously reached in Ref.~\cite{Sung:2019xie} and in contrast to those of Refs.~\cite{1982ApJ...260..868S,Rembiasz:2018lok,Mori:2021pcv}. Our new bounds also strongly constrain the role of particle decays in gamma-ray bursts \cite{Berezhiani:1999qh,Diamond:2021ekg}. Moreover, some explosions must fail to produce the observed population of stellar-mass black holes, and radiative particle decays should not change this picture.

For decay MFPs $\lesssim$\,$10^{11}$\,cm, our constraint becomes yet more restrictive, if extreme LESNe, such as the hydrogen-deficient SN~2008ha with ejecta kinetic energy of only 0.01--0.05\,B \cite{Valenti+2009}, are confirmed as CCSN events of stripped-envelope progenitors.

While our arguments are only based on low explosion energies, referring to energy conservation, yet more sensitive observables could be the very low luminosities of LESNe as well as their light-curve shape and spectral line velocities. Momentum and energy deposition by particle decays in the outer stellar layers far ahead of the slower SN shock might lead to an unusual structure of the early light curve and an anomalous evolution of the photospheric velocities. The subsequent shock collision with decay-driven preceding ejecta could also cause peculiar brightness variations. However, predicting the radiation emission for comparison to observations requires hydrodynamic modeling of the energy deposition and mass ejection, including radiative transfer, which is a task that must be left for future work.

If the initial luminosity rise of a next SN in our own galaxy is observed, a brightness increase already around 100\,s after the measurement of the $\sim$10\,s long neutrino burst of this SN would be a spectacular indication for energy deposition by particle decays in the outer stellar layers. Conversely, the absence of this effect would provide new, more sensitive, constraints.

\textit{Acknowledgments.---}%
AC is supported by the Foreign Postdoctoral Fellowship Program of the Israel Academy of Sciences and Humanities. AC also acknowledges support from the Israel Science Foundation (Grant 1302/19), the US-Israeli BSF (Grant 2018236) and the German-Israeli GIF (Grant I-2524-303.7). 
EV was supported in part by the US\ Department of Energy (DOE) Grant DE-SC0009937. 
HTJ and GR acknowledge support by the German Research Foundation (DFG) through the Collaborative Research Centre ``Neutrinos and Dark Matter in Astro and Particle Physics (NDM),'' Grant SFB-1258, and under Germany’s Excellence Strategy through the Cluster of Excellence ORIGINS EXC-2094-390783311.


\bibliographystyle{bibi}
\bibliography{biblio}

\onecolumngrid
\appendix

\clearpage

\setcounter{equation}{0}
\setcounter{figure}{0}
\setcounter{table}{0}
\setcounter{page}{1}
\makeatletter
\renewcommand{\theequation}{S\arabic{equation}}
\renewcommand{\thefigure}{S\arabic{figure}}
\renewcommand{\thepage}{S\arabic{page}}

\begin{center}
\textbf{\large Supplemental Material\\[0.5ex]
Low-Energy Supernovae Severely Constrain Radiative Particle Decays}
\end{center}

In this Supplemental Material, we provide details about the ALP emission processes as well as a description of how to apply redshift corrections to radiation and massive particles that escape the gravitational potential inside proto-neutron stars.

\bigskip

\twocolumngrid

\section{A.~Primakoff Process} 

Based on the interaction $\mathcal{L}_{a\gamma\gamma}=G_{a\gamma\gamma}a\, {\bf E}\cdot{\bf B}$, the differential cross section for $\gamma+Ze\to Ze+a$, averaged over photon polarizations, is \cite{Raffelt:1985nk}
\begin{equation}\label{eq:Primakoff-differential}
    \frac{d\sigma_{\rm P}}{d\Omega}=\frac{Z^2\alpha G_{a\gamma\gamma}^2}{8\pi}\,
    \frac{|{\bf k}\times{\bf p}|^2}{{\bf q}^4}\,\frac{{\bf q}^2}{{\bf q}^2+k_{\rm s}^2},
\end{equation}
where ${\bf k}$ is the photon momentum, ${\bf p}$ the ALP one, ${\bf q}={\bf p}-{\bf k}$ the momentum transfer, and $k_{\rm s}$ the screening scale in the medium. The target $Ze$ is taken to be at rest and static (no recoil).

A precise evaluation in a SN core is difficult because electrons are relativistic and degenerate, however implying that their contribution both as scattering targets and for screening is small. We follow our earlier discussion \cite{Caputo:2021rux} and neglect electrons, so for screening we use the Debye-H\"uckel scale of the quasi-static ions
\begin{equation}\label{eq:screeningscale}
  \kS^2=\frac{4\pi\alpha \hat n}{T}
\quad\hbox{where}\quad
  \hat{n}=\sum_j Z_j^2 n_j\equiv\hat{Y}n_B,
\end{equation}
which defines the effective charge $\hat{Y}$ per baryon. While in a SN core, the charged particles are typically protons, in some regions small nuclear clusters dominate instead. Following our earlier discussion \cite{Caputo:2021rux}, we use $\hat Y=1-Y_n$ for estimating the Primakoff production and absorption rate in a given SN model.

Photons in the medium suffer nontrivial dispersion that is dominated by the relativistic electron plasma frequency $\wP^2=(4\alpha/3\pi)(\mu_e^2+\pi^2T^2/3)$, where $\mu_e$ is the electron chemical potential. For typical SN conditions this is $\wP\simeq10~\MeV$, to be compared with $T\simeq30\,\MeV$ and a typical photon energy $\omega\simeq 3T\simeq100\,\MeV$. Therefore, we approximate photons as massless particles. Otherwise we should also include a wave-function renormalization factor in Eq.~\eqref{eq:Primakoff-differential} that was approximated as~1.

With $|{\bf k}\times{\bf p}|^2=k^2 p^2\sin^2\theta$ (scattering angle $\theta$) and ${\bf q}^2=p^2+k^2-2pk\cos\theta$, the angular integral in Eq.~\eqref{eq:Primakoff-differential} can be performed explicitly and yields \cite{DiLella:2000dn}
\begin{equation}
    \sigma_{\rm P}=\frac{Z^2\alpha G_{a\gamma\gamma}^2}{2}\,f_{\rm P},
\end{equation}
where
\begin{eqnarray}
    f_{\rm P}&=&\frac{\bigl[(k+p)^2+\kS^2\bigr]\bigl[(k-p)^2+\kS^2\bigr]}{16\kS^2kp}
    \log\frac{(k+p)^2+\kS^2}{(k-p)^2+\kS^2}
    \nonumber\\[1ex]
    &&{}-\frac{(k^2-p^2)^2}{16\kS^2kp}\log\frac{(k+p)^2}{(k-p)^2}-\frac{1}{4}.
\end{eqnarray}
In the limit of massless ALPs and photons where $p=k=\omega$ this is
\begin{equation}
    f_{\rm P}=\frac{1}{4}
    \left[\left(1+\frac{\kS^2}{4\omega^2}\right)\log\left(1+\frac{4\omega^2}{\kS^2}\right)-1\right].
\end{equation}
The cross section depends only logarithmically on the screening scale. In a SN core, $\kS\alt T$ and $\omega\simeq 3T$. Taking $\kS/\omega=1/4$ we find $f_{\rm P}=0.810$ and a proper thermal average $\langle f_{\rm P}\rangle$ varies between about 0.5 and 1 in the crucial SN regions (see Fig.~4 of Ref.~\cite{Caputo:2021rux}).

In the opposite limit of a large ALP mass so that $p\ll\omega$, one finds
\begin{equation}\label{eq:fPexpansion}
    f_{\rm P}=\frac{2\omega^2}{3(\omega^2+\kS^2)}\,v^2+{\cal O}(v^4),
\end{equation}
where $v=p/\omega=\sqrt{1-m_a^2/\omega^2}$ is the ALP velocity.
Therefore, ALP production near threshold is suppressed by a factor $v^2$.

If the ALP mass is the dominant scale for cutting off the Coulomb divergence, i.e., for $m_a\agt30\,\MeV$, we may set $\kS=0$ and~find
\begin{equation}
    f_{\rm P}=\frac{1+v^2}{4v}\log\frac{1+v}{1-v}-\frac{1}{2}=\frac{2}{3}v^2+{\cal O}(v^4).
\end{equation}
Assuming $m_a= 30\,\MeV$ and $\omega=100\,\MeV$ we find $v=0.954$ and $f_{\rm P}=1.38$.

Ignoring degeneracy effects for the charged targets, the energy-loss rate per unit volume of the SN medium is
\begin{eqnarray}
    Q_a&=&\int\frac{2d^3{\bf k}}{(2\pi)^3}\frac{\omega}{e^{\omega/T}-1} \hat n\sigma_{\rm P}
    \nonumber\\
    &=&\hat n\,\alpha G_{a\gamma\gamma}^2\,\frac{\pi^2 T^4}{30}\,\langle f_{\rm P}\rangle,
\end{eqnarray}
where the first factor of 2 counts two photon polarizations and the lower limit of the photon energy integral is $m_a$. The thermally averaged screening factor $\langle f_{\rm P}\rangle$ depends on $\kS$, $m_a$, and $T$. For $m_a\gg T$ we may use $f_{\rm P}=2 v^2/3$ and~find
\begin{equation}\label{eq:fPapprox}
    \langle f_{\rm P}\rangle=20\frac{m_a^2+3m_a T+3T^2}{\pi^4 T^2}\,e^{-m_a/T}.
\end{equation}
For $m_a=0$ this expression is 0.616, whereas for $\kS=30\,\MeV$ the full expression for $m_a=0$ yields 0.734. Therefore, throughout our SN model, Eq.~\eqref{eq:fPapprox} is a reasonable approximation (within a few 10\%) for all $m_a$, not just asymptotically for large masses.

\begin{figure}[ht]
\vskip-6pt
\centering
\includegraphics[width=0.85\columnwidth]{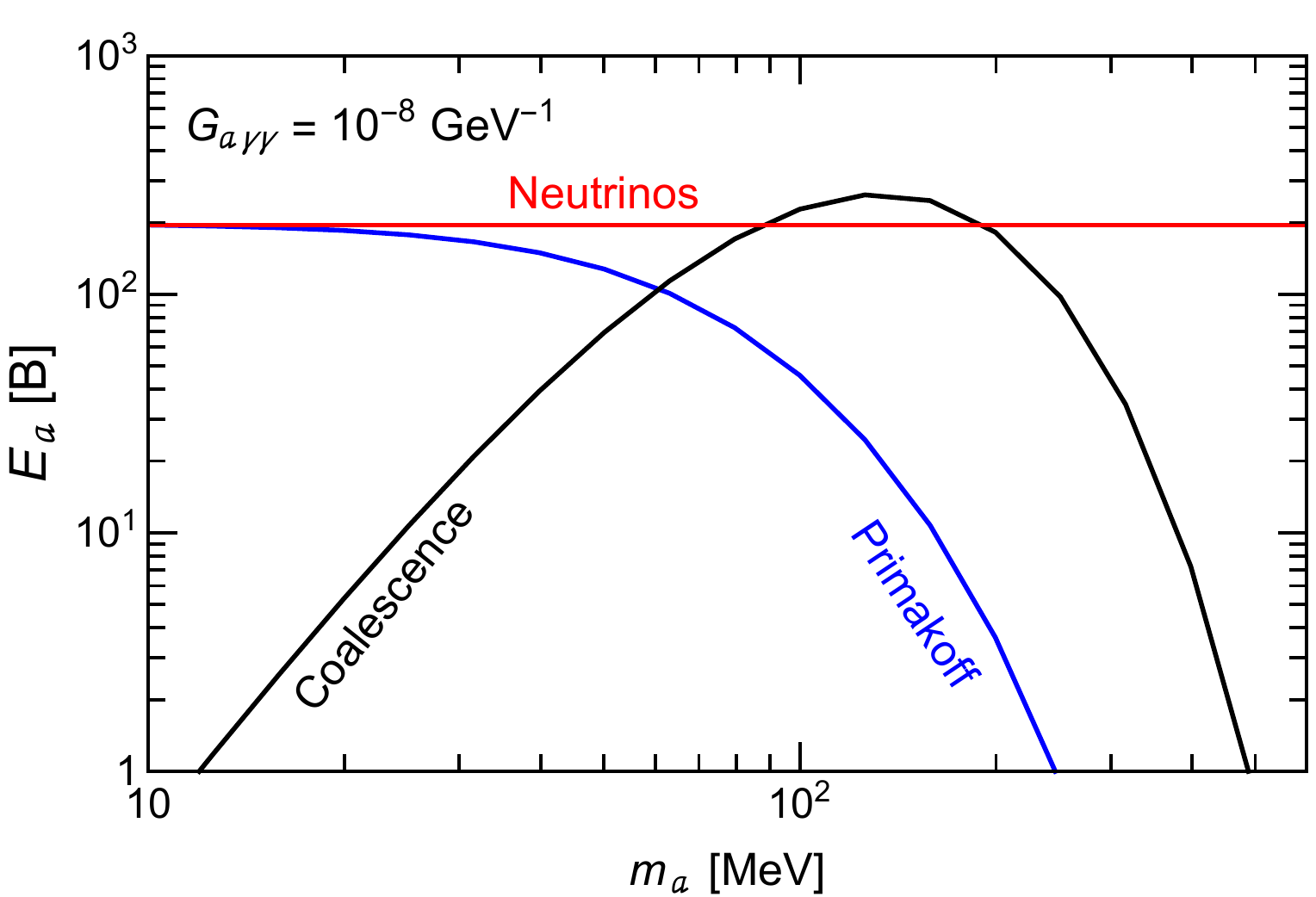}
\vskip-6pt
\caption{Time-integrated energy released in ALPs by our reference SN model for $G_{a\gamma\gamma} =  10^{-8}\,\GeV^{-1}$ by Primakoff emission (blue curve) and $\gamma\gamma$ coalescence (black curve). The horizontal red line is the total energy emitted in neutrinos.}\label{fig:Ene}
\vskip-6pt
\end{figure}

For our reference SN model, we show in Fig.~\ref{fig:Ene} the time-integrated ALP energy emission as a function of $m_a$, based on the full expression. For $m_a\agt 30\,\MeV$ the emission begins being suppressed by $m_a$.

\section{B.~Photon Coalescence} 

To calculate the energy-loss rate from photon coalescence $2\gamma\to a$ we ignore photon dispersion effects and begin with the reverse process of ALP decay. In vacuum it has the well-known rate $\Gamma_{a\to2\gamma}=G_{a\gamma\gamma}^2 m_a^3/64\pi$, in the laboratory frame to be supplemented with the Lorentz factor $m_a/\omega$ with $\omega$ the ALP energy. In the thermal SN environment, the ALP absorption rate finally is
\begin{equation}
    \Gamma_{\rm A}=\frac{G_{a\gamma\gamma}^2 m_a^4}{64\pi\,\omega}\,f_{\rm B},
\end{equation}
where $f_{\rm B}$ accounts for Bose-Einstein stimulation of the final-state photons.

A decay photon with energy $\omega_i$, $i=1$ or 2, is stimulated with $1+(e^{\omega_i/T}-1)^{-1}=(1-e^{-\omega_i/T})^{-1}$, so the overall stimulation factor is
$(1-e^{-\omega_1/T})^{-1}(1-e^{-\omega_2/T})^{-1}$. The two photon energies add up to $\omega_1+\omega_2=\omega$. Each of the decay photons is uniformly distributed on the interval $(\omega-p)/2\leq\omega_i\leq(\omega+p)/2$, where the interval has length $p$. Averaging the stimulation factor over this energy range we find
\begin{eqnarray}\label{eq:fBfull}
    f_{\rm B}&=&\frac{1}{p}\int_{(\omega-p)/2}^{(\omega+p)/2}
    \frac{d\omega_1}{(1-e^{-\omega_1/T})(1-e^{-(\omega-\omega_1)/T})}
    \nonumber\\[1ex]
    &=&\frac{1}{1-e^{-\omega/T}}\frac{2T}{p}
     \log\left[\frac{e^{\frac{\omega+p}{4T}}-e^{-\frac{\omega+p}{4T}}}{e^{\frac{\omega-p}{4T}}-e^{-\frac{\omega-p}{4T}}}\right]\!.
\end{eqnarray}
In the limit $T\to0$ this expression approaches $f_{\rm B}=1$ and we are back to the vacuum decay rate. In the limit of vanishing momentum (axions at rest with $\omega=m_a$) the expression expands as
$f_{\rm B}=(1-e^{-m_a/2T})^{-2}$. Indeed, in this limit each photon has energy $m_a/2$ and is stimulated
by $(1-e^{-m_a/2T})^{-1}$, so the total stimulation is the square of this factor.

By detailed balance, the spontaneous emission rate for an ALP of energy $\omega$ is $\Gamma_{\rm E}=\Gamma_{\rm A}\,e^{-\omega/T}$ for a fixed momentum mode. Therefore, the energy emission rate per unit volume is the phase-space integral
\begin{equation}
    Q_{a}=\int\frac{d^3{\bf p}}{(2\pi)^3}\,\omega\,e^{-\omega/T}\Gamma_{\rm A}
    =\frac{G_{a\gamma\gamma}^2T^3m_a^4}{128\pi^3}\,F(m_a/T),
\end{equation}
where
\begin{equation}
    F(\mu)=\int_{\mu}^\infty dx\,x \sqrt{x^2-\mu^2}\,e^{-x}f_{\rm B}.
\end{equation}
Here $x=\omega/T$ and $\mu=m_a/T$.

Using Maxwell-Boltzmann (MB) statistics, where $f_{\rm B}=1$, this result agrees with Ref.~\cite{DiLella:2000dn} and with what was used in Ref.~\cite{Lucente:2020whw}. In this case the integral can be evaluated explicitly and we find
\begin{equation}
    F(\mu)\big|_{{\rm MB}}=\mu^2\,K_2(\mu),
\end{equation}
where $K_2(\mu)$ is a modified Bessel function of the second kind, in {\sc Mathematica} notation
{\tt BesselK[2,$\mu$]}. For small arguments it is $2/\mu^2$, for large ones $e^{-\mu}\sqrt{\pi/2\mu}$.

Using instead our full expression Eq.~\eqref{eq:fBfull}, we finally obtain
\begin{equation}
    F(\mu)=
    \int_{\mu}^\infty\! dx\,\frac{2x}{e^{x}-1}
    \log\left[\frac{e^{\frac{x+y}{4}}-e^{-\frac{x+y}{4}}}{e^{\frac{x-y}{4}}-e^{-\frac{x-y}{4}}}\right],
\end{equation}
where $x=\omega/T$ and $y=p/T=\sqrt{x^2-\mu^2}$. For $m_a\gg T$, equivalent to $\mu\gg1$, both the full and MB expressions are asymptotically $F(\mu)\to e^{-\mu}\sqrt{\pi \mu^3/2}$.

\begin{figure}[ht]
\centering
\includegraphics[width=1\columnwidth]{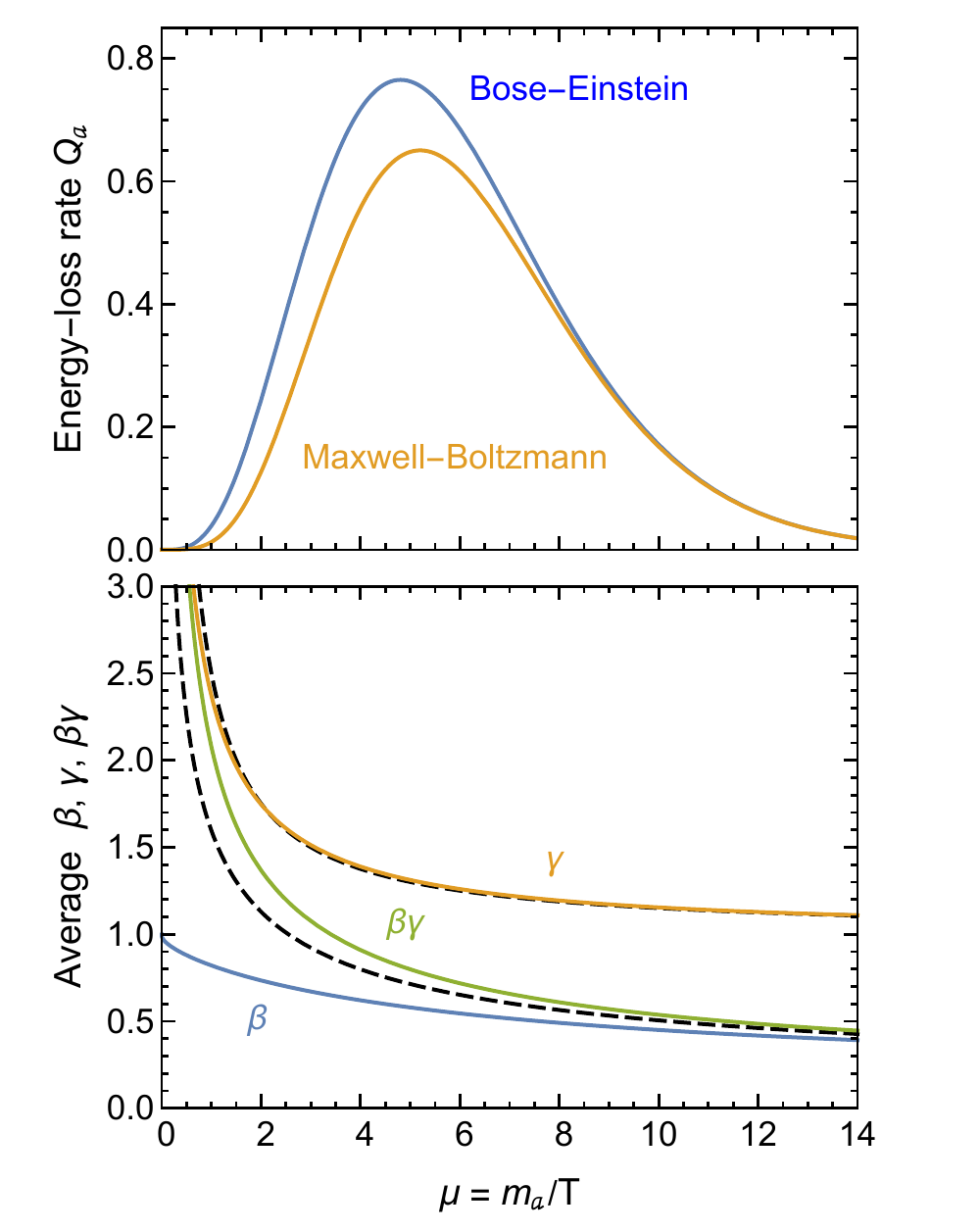}
\caption{\textit{Top:} Energy-loss rate $Q_a$ (in arbitrary units) from photon coalescence, using Bose-Einstein statistics for the photons (blue) or Maxwell-Boltzmann (orange).
\textit{Bottom:} Average ALP velocity $\beta$, energy $\gamma=\omega/m_a$, and $\beta\gamma$. The dashed lines are the large-mass expansions (see text).}\label{fig:MB}
\end{figure}

In Fig.~\ref{fig:MB} we show $Q_a\propto\mu^4 F(\mu)/128$, i.e., the energy-loss rate in arbitrary units, for both Bose-Einstein and Maxwell-Boltzmann photon statistics. While the curves look fairly similar, for $m_a\alt5T$ their ratio deviates significantly from 1. We also see that the maximum of $Q_a$ appears near $m_a=5 T$. For $T=30\,\MeV$ this would be around $m_a=150\,\MeV$. So the photon coalescence process is most efficient at rather large masses compared with the temperature. The same conclusion derives from Fig.~\ref{fig:Ene} where we show the time-integrated
energy-loss rate for our reference SN model.

Based on the ALP energy distribution implied by these results it is straightforward to compute the average of the ALP velocity $\beta$ and energy (or rather Lorentz factor) $\gamma=\omega/m_a$. The quantity $\beta\gamma$ is also of interest for the average mean free path. We show these averages in the bottom panel of Fig.~\ref{fig:MB}. Even though in the range of interest $m_a\gg T$, the average velocity of the emitted ALPs is not very small and intermediate between relativistic and nonrelativistic.

Based on the analytic Maxwell-Boltzmann case, that is exact for large $\mu$, we find the large-$\mu$ results $\langle\beta\rangle=\sqrt{8/\pi\mu}$, $\langle\gamma\rangle=1+3/2\mu$, and $\langle\beta\gamma\rangle=\sqrt{8/\pi\mu}$, shown as dashed lines.

\section{C.~Relativistic corrections and gravitational redshift}

As already explained in Ref.~\cite{Caputo:2021rux}, the Garching SN model~\cite{Bollig:2020xdr} includes general-relativistic effects in an approximate manner (see \cite{Rampp:2002bq}), which for us means that the spatial coordinates should be interpreted as in flat space and the time coordinate is the one of a distant observer. Nevertheless, once an ALP is emitted, it will suffer a gravitational redshift before reaching infinity. This effect is encoded  in the ``gravitational lapse'' factor listed at every radius of the simulation output, which means that the energy of the particle at infinity is $E_{\infty} = E_{\rm loc} \times \rm lapse$. We stress that this rescaling applies to
\textit{both} massless and massive particles. In a gravitational background with given spacetime metric, this redshift effect can be derived from the equations for the free motion of a test particle along geodesics, specifically from the temporal component of the Euler-Lagrange equations. For example, working in geometrized units with $c = G = 1$, in a Schwarzschild geometry exterior to a gravitating mass $M$, one finds $\mathrm{lapse} =[1-2M/r]^{1/2}$, as discussed, for example, in Section 12.4 of~\cite{Shapiro:1983du}. However, the situation in the interior of the gravitating object is more complicated and the tabulated lapse factor includes contributions from the pressure and energy of the stellar medium and of neutrinos besides the leading mass term. For a detailed description of the formalism to treat general relativistic effects in the Garching models, we point the reader to the original references~\cite{Rampp:2002bq,1979ApJ...232..558V}.

Apart from the particle energy, also the rate of emission requires a redshift correction because of the distinction between local time and observer time, which causes a time-stretching of the luminosity measured outside of the strong gravitational field. Therefore, when computing the contributions from local emission to the luminosity arriving at large distances, one needs to include a total factor $(\mathrm{lapse})^2$.  Moreover, the physical properties of the star are given in the comoving frame of the emitting medium. This causes also a Doppler shift effect $\propto$\,$(1 + 2 v_r)$, where $v_{\rm r}$ is the radial velocity (positive or negative) of the medium, which, however, is always very small in the PNS, $|v_r| \ll 1$. Summing up, the total (energy integrated) luminosity of ALPs produced in the PNS and observed at infinity (i.e., very far away from the source as shown in Fig.~\ref{fig:LuminosityProfile}) reads
\begin{equation}
L_\infty = \int_0^{\RNS} dR\, 4\pi R^2 Q_a \cdot (\mathrm{lapse})^2(1+2 v_{\rm r})\,,
\end{equation}
where $Q_a$ denotes the energy emission rate of particles $a$ as measured in the local frames of the PNS matter. Here, we assume a flat space (consistent with the outputs from the pseudo-Newtonian hydrodynamic models used; \cite{Rampp:2002bq}), steady-state emission of the particles (i.e., slow changes of the particle production with time), and a hydrostatic PNS structure ($|v_r| \ll c_\mathrm{s}$). Moreover, because of the very slow evolution of the cooling PNS, changes of the PNS properties on time-of-flight timescales are ignored (saving interpolation of simulation outputs in time).

This recipe is valid for both massive and massless particles, however massive particles need extra care. In fact, locally produced particles must have a kinetic energy that can overcome the gravitational potential in order to escape to infinity. Therefore the local energy of emission needs to satisfy $E_{\rm local} > m_a/ \rm lapse$, where $m_a$ is the ALP rest mass. In the aforementioned Schwarzschild case this would reduce to the familiar condition $E_{\rm loc} > m_a'= m_a / \sqrt{1- 2 M/R} \simeq m_a + M m_a/R$, where in the last step we assumed gravity to be weak.

\end{document}